\documentclass[10pt,conference]{IEEEtran}
\IEEEoverridecommandlockouts
\usepackage{cite}
\usepackage{amsmath,amssymb,amsfonts}
\usepackage{algorithmic}
\usepackage{graphicx}
\usepackage{textcomp}
\usepackage{xcolor}
\usepackage{url}
\usepackage{verbatim}
\usepackage{subfigure}
\usepackage{harpoon}
\usepackage{multirow}
\usepackage{bm}

\usepackage[numbers,sort&compress]{natbib}

\def\BibTeX{{\rm B\kern-.05em{\sc i\kern-.025em b}\kern-.08em
		T\kern-.1667em\lower.7ex\hbox{E}\kern-.125emX}}
\begin{document}
	
\title{No-Reference Light Field Image Quality Assessment Based on Micro-Lens Image\\
\thanks{* Equal contribution}
}
\author{
\IEEEauthorblockN{Ziyuan Luo*, Wei Zhou*, Likun Shi, and Zhibo Chen}
\IEEEauthorblockA{\textit{CAS Key Laboratory of Technology in Geo-spatial Information Processing and Application System}\\
\textit{University of Science and Technology of China, Hefei 230027, China} \\
chenzhibo@ustc.edu.cn}
}
\maketitle

\begin{abstract}
Light field image quality assessment (LF-IQA) plays a significant role due to its guidance to Light Field (LF) contents acquisition, processing and application. The LF can be represented as 4-D signal, and its quality depends on both angular consistency and spatial quality. However, few existing LF-IQA methods concentrate on effects caused by angular inconsistency. Especially, no-reference methods lack effective utilization of 2-D angular information. In this paper, we focus on measuring the 2-D angular consistency for LF-IQA. The Micro-Lens Image (MLI) refers to the angular domain of the LF image, which can simultaneously record the angular information in both horizontal and vertical directions.
Since the MLI contains 2-D angular information, we propose a No-Reference Light Field image Quality assessment model based on MLI (LF-QMLI). Specifically, we first utilize Global Entropy Distribution (GED) and Uniform Local Binary Pattern descriptor (ULBP) to extract features from the MLI, and then pool them together to measure angular consistency. In addition, the information entropy of Sub-Aperture Image (SAI) is adopted to measure spatial quality. Extensive experimental results
show that LF-QMLI achieves the state-of-the-art performance.
\end{abstract}


\section{\textbf{Introduction}}
\label{sec:intro}
As a representative of the attractive technique for immersive multimedia data, Light Field (LF) images have attracted widespread attention~\cite{wu2017light}. Unlike traditional 2D images, LF images can record radiance information in both spatial and angular dimensions~\cite{levoy2009recording}, leading to better immersive experience. In order to provide satisfactory viewing quality of experience (QoE), light field image quality assessment (LF-IQA) plays a crucial role in LF contents acquisition, processing and application.

\begin{figure}
	\centering
	\includegraphics[width=\linewidth]{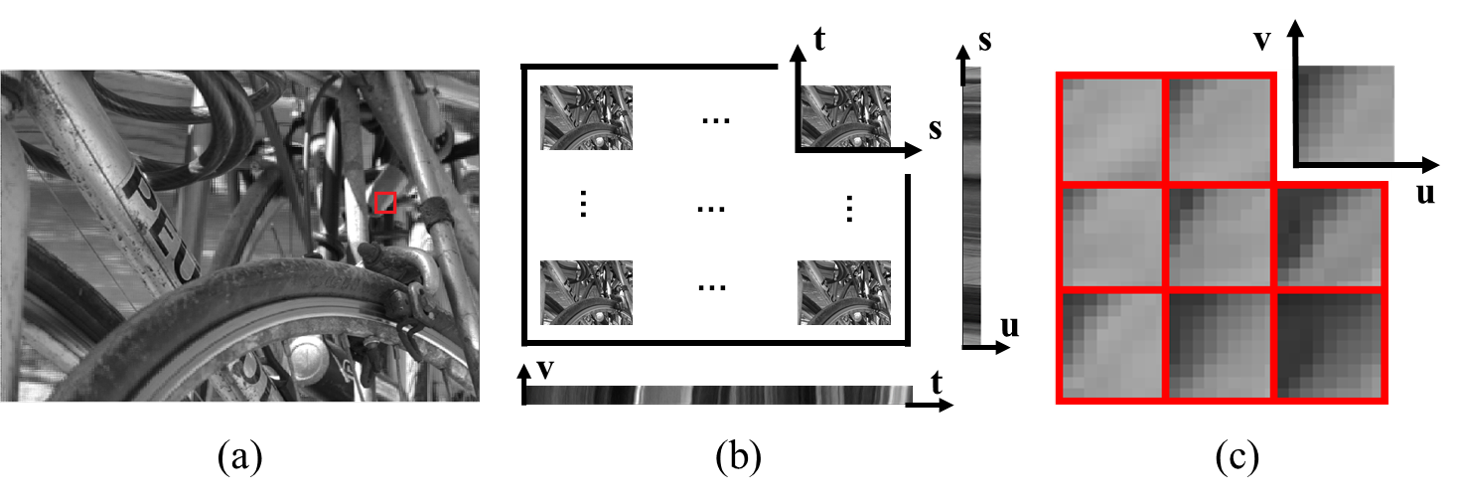}
	\caption{LF image with different formats. (a) Lenset image captured by Lytro Illum; (b) SAI array with EPIs in the bottom and right; (c) 9 MLIs in the red bounding box of (a) at high magnification.}
	\label{fig:intro}
\end{figure}

The LF image is a 4-D signal containing spatial and angular information. In Fig.~\ref{fig:intro}, we show the LF image with different formats. Fig.~\ref{fig:intro}(a) shows the LF image captured by a lenslet LF camera called Lytro Illum~\cite{ng2005light}.
The parameters $u$ and $v$ refer to angular dimensions while $s$ and $t$ represent spatial dimensions. We can obtain the Sub-Aperture Image (SAI) by fixing $u$ and $v$~\cite{van2018light}, while the Micro-Lens Image (MLI) is through fixing $s$ and $t$~\cite{cho2013modeling}, as shown in Fig.~\ref{fig:intro}(b-c). The bottom and right of Fig.~\ref{fig:intro}(b) are Epipolar-Plane Images (EPIs)~\cite{wu2017light}, which are produced by fixing ($u$,$s$) and ($v$,$t$). The SAI only contains spatial information of the LF image, while the EPI include both spatial and angular dimensions. However, the EPI only contains horizontal or vertical angular direction. Unlike the SAI and EPI, the MLI includes 2-D angular information. On account of the 4-D characteristic, the perceptual quality of LF image mainly depends on spatio-angular resolution, angular consistency and spatial quality~\cite{wu2017light}. Concretely, spatio-angular resolution refers to the LF image resolution (i.e. the values of $u$,$v$,$s$ and $t$). Angular consistency measures the visual coherence of LF images while spatial quality indicates the SAI quality. Since spatio-angular resolution is an inherent factor of the LF image, we consider the effects of angular consistency and spatial quality in this paper.

Although the subjective evaluation of LF-IQA ~\cite{viola2016objective, kiran2017towards, shi2018perceptual} is precise and reliable, it is resource and time-consuming. Therefore, an effective objective LF-IQA model is urgently required. In general, image quality assessment (IQA) methods can be classified into three categories: full-reference (FR), reduced-reference (RR) and no-reference (NR)~\cite{zhou2016binocular}. Among FR methods, intact information of original images is needed. Structure similarity between original and distorted images is measured in structural similarity index (SSIM)~\cite{SSIM}, with several variants, e.g. MS-SSIM~\cite{MSSIM} and FSIM~\cite{FSIM}. MP-PSNR Full~\cite{MP_F} and MP-PSNR Reduc~\cite{MP_R} based on Morphological pyramid decomposition are proposed to evaluate the multi-view image quality. 
RR methods only require part of information from original images. NR methods only utilize distorted images, which can be applied to most applications where reference images are hardly available, e.g. Mittal $et \ al.$~\cite{BRI} uses scene statistics in spatial domain and binocular fusion and rivalry are concerned in BSVQE~\cite{BSVQE}. 

There exist only a few objective LF-IQA models. Fang $et \ al.$~\cite{fang2018light} proposes a FR LF-IQA method to compute the gradient magnitude similarity between original and distorted EPIs. Paudyal $et \ al.$~\cite{LFIQM} predicts the LF image quality with the structure similarity between original depth map and distorted depth map. However, neither of the methods consider the spatial quality degradation on the SAI. In addition, the EPIs only contain horizontal or vertical angular dimension, leading to insufficient measurement of angular consistency for the LF image
applications. Therefore, a LF-IQA method that considers spatial quality and 2-D angular consistency is necessary for practical application.

In this paper, we propose a novel NR Light Field image Quality assessment model based on Micro-Lens Image (LF-QMLI) to evaluate both of angular consistency and spatial quality for LF images. As shown in Fig.~\ref{fig:intro}(c), each pixel in the MLI comes from the same point in spatial domain, but is captured in various directions. Hence, there exists quite strong dependence between MLI pixels for 2-D angular consistency. To our best knowledge, we are the first to utilize MLI to evaluate angular consistency of LF images.

In this work, we first obtain the MLI by fixing $s$ and $t$, while the SAI is generated through fixing $u$ and $v$. Second,  the Global Entropy Distribution (GED) and Uniform Local Binary Pattern descriptor (ULBP) are proposed to measure the angular consistency on each MLI, and then we utilize content pooling. Third, the information entropy of SAI is utilized to evaluate spatial quality. Finally, we train a regression model that predicts the perceptual quality of distorted LF images. Our experimental results demonstrate that proposed model LF-QMLI achieves the state-of-the-art performance.

The rest of the paper is organized as follows: Section~\ref{method} introduces our LF-QMLI method. Experimental results are shown in Section~\ref{experiment} and we conclude the paper in Section~\ref{conclusion}.

\begin{figure}
	\centering
	\includegraphics[width=0.9\linewidth]{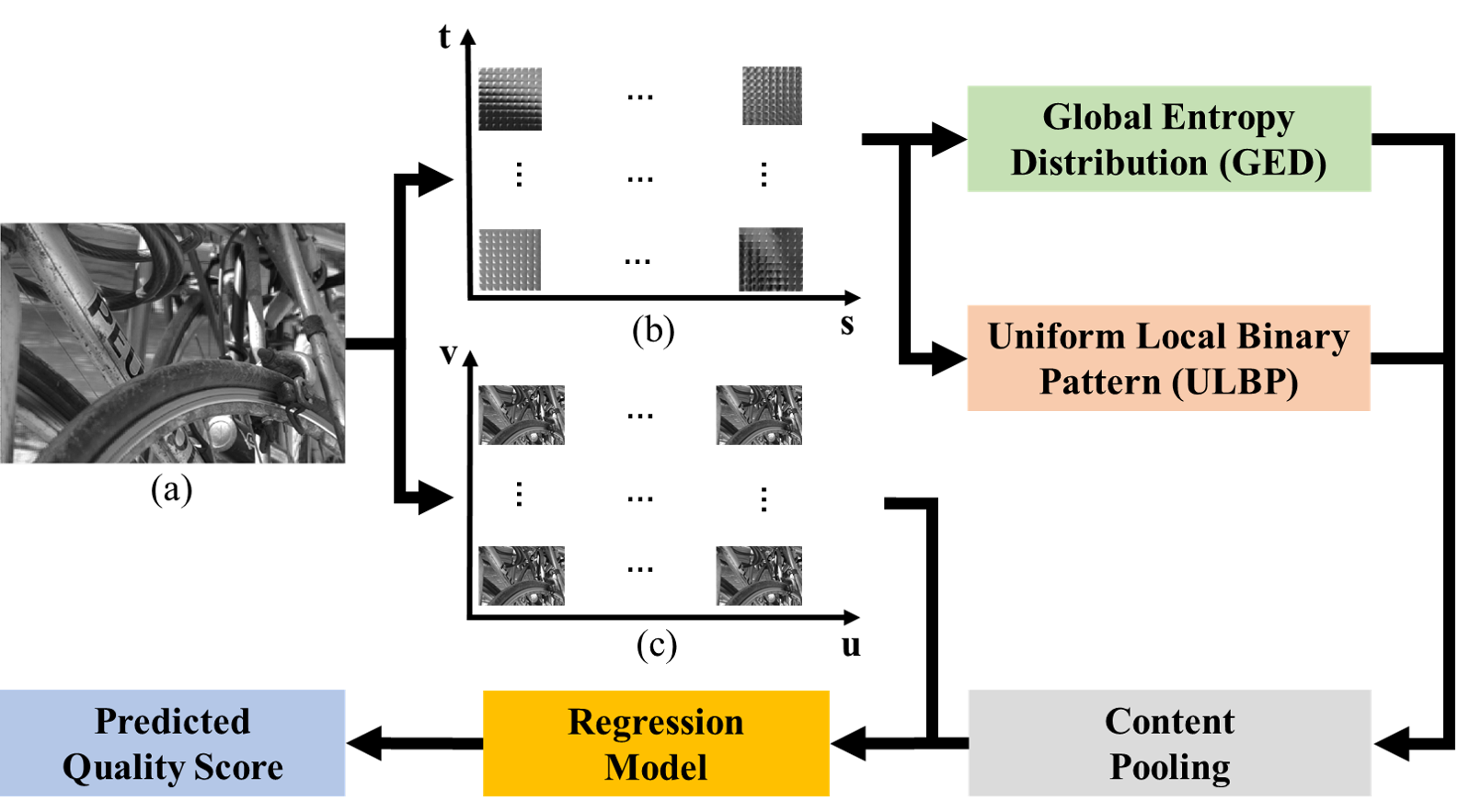}
	\caption{Flow diagram of proposed LF-QMLI model. (a) LF image (lenslet format); (b) MLI Array; (c) SAI Array.}
	\label{fig:flowdiagram}
\end{figure}

\section{\textbf{Proposed Method}}
\label{method}
The flow diagram of our proposed LF-QMLI model is shown in Fig.~\ref{fig:flowdiagram}. We first convert LF images from lenslet format into MLI and SAI arrays. Since the entropy can measure the angular dependence between adjacent pixels~\cite{liu2014no}, the GED is adopted on MLI to measure angular consistency. As the textural variation shown in Fig.~\ref{fig:MLIS}, the ULBP is selected to measure the textural features for original and distorted MLIs. In addition, the information entropy of SAI is utilized to measure spatial quality~\cite{hu2008constructing}.
After content pooling, a regression model is used to predict the perceptual quality of LF images.

\subsection{\textbf{Angular Consistency Based on MLI}}
The distortion caused by angular inconsistency affects LF image quality. A LF camera captures the same object in spatial domain with various angles of view, and the MLI is composed of light rays from both horizontal and vertical directions, leading to a 2-D angular domain.       
\subsubsection{\textbf{Global Entropy Distribution (GED)}}
\label{sec:GED}
In previous works, information entropy is proved as an efficient method to measure spatial image quality~\cite{sponring1996entropy}. However, without considering angular dimension, it cannot work well in LF images.

\begin{figure}
	\centering
	\subfigure[]{
		\begin{minipage}[t]{0.17\linewidth}
			\centering
			\includegraphics[width=\linewidth]{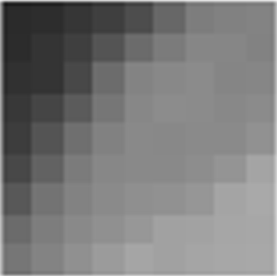}
			\label{subfig:Ori}
	\end{minipage}}
	\subfigure[]{
		\begin{minipage}[t]{0.17\linewidth}
			\centering
			\includegraphics[width=\linewidth]{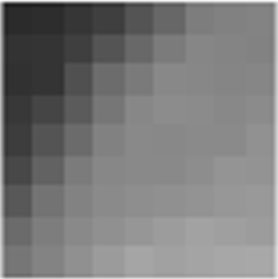}
			\label{subfig:LN20}
	\end{minipage}}	
	\subfigure[]{
		\begin{minipage}[t]{0.17\linewidth}
			\centering
			\includegraphics[width=\linewidth]{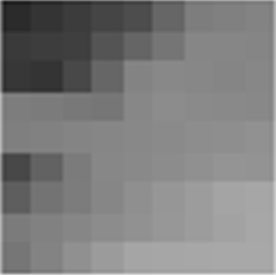}
			\label{subfig:LN50}
	\end{minipage}}
	\subfigure[]{
		\begin{minipage}[t]{0.17\linewidth}
			\centering
			\includegraphics[width=\linewidth]{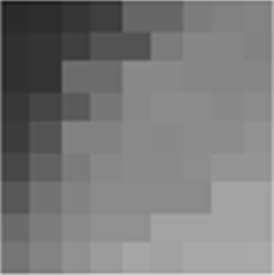}
			\label{subfig:NN20}
	\end{minipage}}
	\subfigure[]{
		\begin{minipage}[t]{0.17\linewidth}
			\centering
			\includegraphics[width=\linewidth]{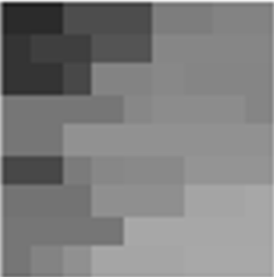}
			\label{subfig:NN50}
	\end{minipage}}
	\caption{MLIs with various types of distortion in different quality levels. Note that distortions in higher level represent worse visual quality. (a) Original Image; (b) LN lv.2 Distorted Image; (c) LN lv.5 Distorted Image; (d) NN lv.2 Distorted Image; (e) NN lv.5 Distorted Image.}
	\label{fig:MLIS}
\end{figure}

The entropy of undistorted images possesses certain statistical properties, owing to the dependence between adjacent pixels~\cite{liu2014no}. As shown in Fig.~\ref{subfig:Ori}, since the variation between pixels is piecewise linear~\cite{heber2013variational}, the MLI without angular distortion is regular and gradually varied. With increased distortion, the dependence between adjacent pixels is destroyed, leading to the change of global entropy. In Fig.~\ref{fig:MLIS}(b)-(e), the distortion caused by angular inconsistency primarily affects the MLI entropy. 

Our proposed GED includes global image entropy distribution and global frequency entropy distribution of MLI.

The image entropy is
\begin{equation}
	E_{I}=-\sum_{x}P_{x}log_{2}P_{x},
\end{equation}
where $x$ is the pixel value within a MLI, ranging from 0 to 255, with empirical probability density $P_{x}$.

The frequency entropy is
\begin{equation}
	E_{F}=-\sum_{i}\sum_{j}P_{i,j}log_{2}P_{i,j},
\end{equation}
where $P_{i,j}$ is the value of probability map locating at $(i,j)$ in the DCT domain of MLI.

Finally, the MLI global entropy $E_{MLI}$ includes image entropy and frequency entropy.
\begin{equation}
	E_{MLI}=\{E_{I},E_{F}\}.
\end{equation}

\begin{table}[htp]
	\centering
	\caption{E$_{MLI}$ on various angular distortion}
	\label{tab:EMLI}
	\renewcommand\arraystretch{1.35}
	\begin{tabular}{c|c|c|c|c|c}
		\hline
		\bf E$_{MLI}$ & \bf Ori. Img & \bf LN lv.2 & \bf LN lv.5 & \bf NN lv.2 & \bf NN lv.5 \\
		\hline
		\bf IE & 5.5772 & 5.4971 & 5.3812 & 5.1807 & 4.3343 \\
		\hline 
		\bf FE & 2.0344 & 2.0543 & 3.0031 & 2.2356 & 3.3460 \\
		\hline
	\end{tabular}
\end{table}

We conducted validation experiments on various levels of distortion caused by angular inconsistency. We considered 2 kinds of angular distortion \{linear interpolation (LN), nearest neighbor interpolation (NN)\} and 2 levels of distortion \{Level2, Level5\}~\cite{shi2018perceptual}. The higher level represents higher distortion. The image entropy (IE) and frequency entropy (FE) of 5 MLIs (Fig.~\ref{fig:MLIS}) were computed to demonstrate the validity of $E_{MLI}$. It is shown in Table~\ref{tab:EMLI} that the angular distortion can affect $E_{MLI}$ in a conspicuous and predictable way. The loss of image details is caused by the distortion, leading to image entropy reduction. Generally, the angular distortion of higher level obtains smaller IE and greater FE. On the same distortion level, NN destroys angular consistency more acutely than LN, which is verified in subjective experiments~\cite{shi2018perceptual}. 

However, there exist a large number of MLIs in the LF image, so we utilize the GED considering all MLIs in the LF image. We will propose our Content Pooling method in section~\ref{sec:pooling}.

\subsubsection{\textbf{Uniform Local Binary Pattern (ULBP)}}
Although image entropy and frequency entropy can measure the distortion caused by angular inconsistency, $E_{MLI}$ is the global characteristics of the whole MLI. As shown in Fig.~\ref{fig:MLIS}, the increased angular distortion changes the local texture of the MLI. Thus, we utilize ULBP to measure local textural features in MLI.

Local binary patterns (LBP) has been proved as an efficient operator to extract local distribution information~\cite{ojala1994performance, ojala1996comparative}. LBP is a very simple but efficient texture descriptor, with rotation invariance, position invariance and robustness under various illumination.  Since LBP can efficiently represent local distributions of joint pixels, we adopt modified ULBP descriptor to describe inconsistency of local adjacent pixels.

The LBP operator can be given as~\cite{ojala2002multiresolution}
\begin{equation}
LBP_{P,R}=\sum_{p=0}^{P-1}s(g_{p}-g_{c})2^{p}.
\end{equation}

Here we set a neighbor of $P=4$ members on a circle of radius $R=1$. $g_{c}$ is the gray value of central pixel, while $g_{p}$ is the gray value of the neighbor pixel. $s$ represents the sign function. 

To reduce the number of pattern types, we utilize modified Uniform LBP operator on each MLI$_{i}$. There exist $P+2$ types of uniform pattern class without regard to upright property. Then we combine the Probability of every Type of binary patterns ($PT$) into vector $\bm{PT}_{i}$.  
\begin{equation}
\bm{PT}_{i}=\{\underbrace{PT_{1},PT_{2},\cdots,PT_{n}}_{P+2}\}.
\end{equation}

\subsubsection{\textbf{Content Pooling}}
\label{sec:pooling}
In order to evaluate LF image quality, we pool the characteristics $E_{MLI}$ and $\bm{PT}_{i}$ of each MLI together into LF image angular features.

The method of pooling used on $E_{MLI}$ is percentile pooling~\cite{moorthy2009visual}: central elements of $E_{MLI}$ are extracted while extremely large or small elements are neglected. Our experimental results verify that percentile pooling improves our proposed model.

\begin{figure}
	\centering
	\includegraphics[width=\linewidth]{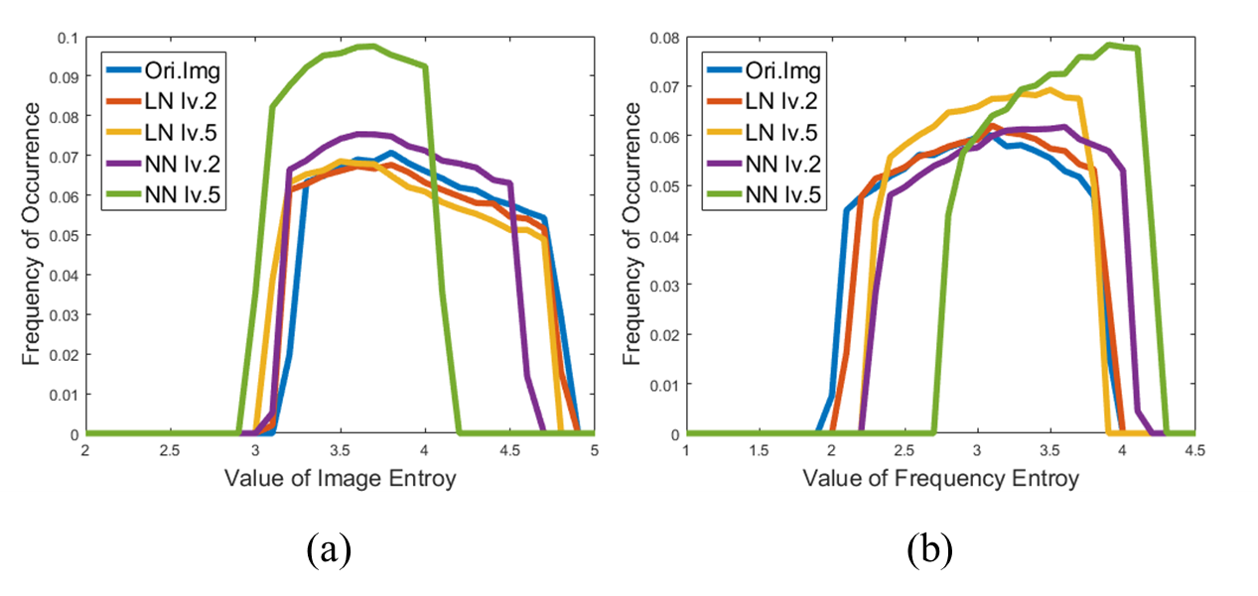}
	\caption{Histograms of GED for different types and levels of angular distortion. (a) IE histogram; (b) FE histogram.}
	\label{fig:GEDhistogram}
\end{figure}

We reserve 60\% of central elements of $E_{MLI}$ here and show the global distribution histogram of IE and FE in Fig.~\ref{fig:GEDhistogram}. In accord with our analysis of Table~\ref{tab:EMLI}, with the increase of angular distortion, the histogram of IE has a left shift, while the histogram of FE has a right shift. The higher level the distortion is, the larger the shift will become and the steeper the curve will appear.

The mean values and skew values of IE and FE distribution histogram are selected as the GED features.
\begin{equation}
f_{GED}=\{ mean(IE), skew(IE), mean(FE), skew(FE) \}.
\end{equation}

The angular inconsistency of distorted LF images appears obviously at the edge between the foreground and background, while it appears inconspicuously in the background or mild areas. In these mild MLIs, less information is contained and the ULBP features might be misleading. Therefore, in order to exclude the MLIs in mild areas, we introduce a selector on $\bm{PT}_{i}$. The ULBP features of a whole LF image are exacted as:
\begin{equation}
f_{ULBP}=avg\{\bm{PT}_{i}\} \quad \forall. i \quad if \quad R(i)>threshold,
\end{equation}
where $R(i)=Max(MLI_{i})-Min(MLI_{i})$, the Range of a single MLI$_{i}$. $threshold$ is selected as gray value 20 here.  

\subsection{\textbf{Spatial Quality}}
The spatial quality plays an important part in LF image perceptual quality. Specifically, we utilize information entropy distribution of SAI to measure the changes in spatial quality. In the undistorted SAI, there exists the spatial dependence between adjacent pixels~\cite{liu2014no}. With increased spatial distortion such as compression, the dependence is destroyed, resulting in changes of information entropy.

We divide the SAI into $8 \times 8$ blocks, then compute SAI Image Entropy (SIE) and SAI Frequency Entropy (SFE) of each block. Finally, pooling all the blocks together like what is shown in section~\ref{sec:pooling} and obtain Spatial Quality features.
\begin{equation}
\begin{aligned}
f_{SQ}=\{ \ mean(SIE),skew(SIE), \\
	mean(SFE),skew(SFE) \ \}. 
\end{aligned}
\end{equation}

\section{\textbf{Experiment results}}
\label{experiment}

\subsection{\textbf{Light Field Image Databases}}
To test the performance of our proposed LF-QMLI model, comparison experiments were conducted on Win5-LID~\cite{shi2018perceptual} and VALID~\cite{viola2018valid} databases. In Win5-LID database, there exist 220 distorted LF images with various distortion types and levels. The distortion types consist of JPEG2000, HEVC, linear interpolation (LN), nearest neighbor interpolation (NN) and two CNN models. The overall Mean Opinion Score (MOS) value is provided for each LF image.   

VALID database includes 5 reference LF images captured by Lytro Illum with a number of distorted LF images caused by several types of compression methods. In this experiment, we utilize several distorted LF images obtained through the interactive methodology, including 40 LF images with two types of distortion: HEVC and VP9.

\begin{table}[htbp]
	\centering
	\caption{Performance Comparison}
	\label{tab:performance}
	\renewcommand\arraystretch{1.3}
	\setlength{\tabcolsep}{0.5mm}{
		\begin{tabular}{c|ccc|ccc}
			\hline
			& \multicolumn{3}{|c|}{\bf Win5-LID} & \multicolumn{3}{|c}{\bf VALID} \\
			\hline
			\bf Metrics & \bf SROCC & \bf LCC & \bf RMSE & \bf SROCC & \bf LCC & \bf RMSE \\
			\hline
			\bf PSNR               & 0.6026 & 0.6189 & 0.8031 & 0.9620 & 0.9681 & 0.3352 \\
			\bf SSIM~\cite{SSIM}   & 0.7346 & 0.7596 & 0.6650 & 0.9576 & 0.9573 & 0.3868 \\
			\bf MS-SSIM~\cite{MSSIM} & 0.8266 & 0.8388 & 0.5566 & 0.9593 & 0.9658 & 0.3473 \\
			\bf FSIM~\cite{FSIM}   & 0.8233 & 0.8318 & 0.5675 & 0.9695 & 0.9798 & 0.2678 \\
			\bf IWSSIM~\cite{IW}   & 0.8352 & 0.8485 & 0.5492 & 0.9674 & 0.9764 & 0.2892 \\
			\bf IFC~\cite{IFC}     & 0.5028 & 0.5393 & 0.8611 & 0.9693 & \bf 0.9909 & \bf 0.1800 \\
			\bf VIF~\cite{VIF}     & 0.6665 & 0.7032 & 0.7270 & \bf 0.9749 & 0.9870 & 0.2150 \\
			\bf NQM~\cite{NQM}     & 0.6508 & 0.6940 & 0.7362 & 0.9055 & 0.9194 & 0.5266 \\
			\bf VSNR~\cite{VSNR}   & 0.3961 & 0.5050 & 0.8826 & 0.9359 & 0.9324 & 0.4838 \\
			\hline
			\bf BRISQUE~\cite{BRI} & 0.6687 & 0.7510 & 0.5619 & 0.9222 & 0.9849 & 0.2017 \\
			\bf NIQE~\cite{NIQE}   & 0.2086 & 0.2645 & 0.9861 & 0.8636 & 0.9524 & 0.4080 \\
			\bf FRIQUEE~\cite{FRI} & 0.6328 & 0.7213 & 0.5767 & 0.9157 & 0.9836 & 0.2160 \\
			\hline
			\bf Chen~\cite{Chen}   & 0.5269 & 0.6070 & 0.8126 & 0.9642 & 0.9738 & 0.3046 \\
			\hline
			\bf SINQ~\cite{SINQ}   & 0.8029 & 0.8362 & 0.5124 & 0.9222 & 0.9849 & 0.2070 \\
			\bf BSVQE~\cite{BSVQE} & 0.8179 & 0.8425 & 0.4801 & 0.9222 & 0.9814 & 0.2180 \\
			\hline
			\bf MP-PSNR Full~\cite{MP_F}  & 0.5335 & 0.4766 & 0.8989 & 0.9730 & 0.9852 & 0.2291 \\
			\bf MP-PSNR Reduc~\cite{MP_R} & 0.5374 & 0.4765 & 0.8989 & 0.9744 & 0.9859 & 0.2237 \\
			\bf MW-PSNR Full~\cite{MW}  & 0.5147 & 0.4758 & 0.8993 & 0.9597 & 0.9677 & 0.3376 \\
			\bf MW-PSNR Reduc~\cite{MW} & 0.5326 & 0.4766 & 0.8989 & 0.9648 & 0.9751 & 0.2970 \\
			\bf 3DSwIM~\cite{DSwIM}      & 0.4320 & 0.5262 & 0.8695 & 0.7950 & 0.7876 & 0.8248 \\
			\hline
			\bf APT~\cite{APT}     & 0.3058 & 0.4087 & 0.9332 & 0.4699 & 0.6452 & 1.0228 \\
			\hline
			\bf LF-IQM~\cite{LFIQM}& 0.4503 & 0.4763 & 0.8991 & 0.3934 & 0.5001 & 1.1593 \\
			\hline
			\bf LF-QMLI & \bf 0.8802 & \bf 0.9038 & \bf 0.4147 & 0.9286 & 0.9683 & 0.2791\\
			\hline
		\end{tabular}
	}
\end{table} 

\subsection{\textbf{Comparison with Previous Objective Metrics}}

We conducted comparison experiments between our proposed model and several FR, RR and NR metrics, including nine 2D-FR metrics~\cite{SSIM,MSSIM,FSIM,IW,IFC,VIF,NQM,VSNR}, three 2D-NR metrics~\cite{BRI,NIQE,FRI}, one 3D-FR metric~\cite{Chen}, two 3D-NR metrics~\cite{SINQ,BSVQE}, five Multi-views FR metrics~\cite{MP_F,MP_R,MW,DSwIM}, one Multi-views NR metric~\cite{APT} and one LF-RR metric~\cite{LFIQM}. Three evaluation criteria are selected to measure the correlation between MOS and predicted results, consisting of Spearman Rank Order Correlation Coefficient (SROCC), Linear Correlation Coefficient (LCC) and Root Mean Squared Error (RMSE). The SROCC measures the monotonicity while LCC evaluates the linear relationship between predicted score and MOS. The RMSE computes the deviation of prediction. The better consistency with human perception is reflected in SROCC and LCC closing to 1 as well as RMSE closing to 0.

Then, we use SVR for regression~\cite{smola2004tutorial}. LIBSVM package~\cite{chang2011libsvm} is utilized to implement the SVR, which uses a radial basis function (RBF) kernel. We randomly select 80\% of the database as the training set while the remaining 20\% constitute the test set. The median of correlation coefficients across 1000 random trails were regarded as the final results. 

The results of all metrics are shown in Table~\ref{tab:performance}. Here we can find that almost all 2D-FR metrics perform well on VALID database and LF-QMLI is competitive in NR metrics. This situation may be caused by its limited types of distortion. VALID only introduces two compression distortions, which destroy the spatial quality of LF images but do not take angular consistency into consideration, so previous 2D-FR metrics can almost excellently measure the distortion. The results show that VALID is not challenging for quality assessment of LF images. Therefore, we mainly analyze how our proposed model LF-QMLI performs on Win5-LID database.

On Win5-LID database, LF-QMLI outperforms all previous metrics. In general, the existing 2D and 3D metrics only consider the degradation of spatial quality, ignoring the degradation of angular consistency. Although multi-view metrics can measure the angular distortion, they do not take the compression distortion and similar spatial distortions into account. Therefore, we can reach a conclusion that the proposed model LF-QMLI can evaluate both angular consistency and spatial quality. 

\subsection{\textbf{Ablation Study}}
In order to verify the validity of our proposed MLI-based model, we conducted an ablation study on Win5-LID database and the results are demonstrated in TABLE~\ref{tab:ablation}. The features extracted from MLI $f_{MLI}$ can obviously improve the model performance. One possible reason is that the $f_{MLI}$ provides the measurement of the 2-D angular consistency.

\begin{table}[htbp]
	\centering
	\caption{Ablation Study}
	\label{tab:ablation}
	\renewcommand\arraystretch{1.2}
	\begin{tabular}{c|ccc}
		\hline
		& SROCC & LCC & RMSE \\
		\hline
		\bf Model-f$ \bf _{MLI}$ & 0.6927 & 0.7890 & 0.5504 \\
		\bf Model & 0.8802 & 0.9038 & 0.4147 \\
		\hline
	\end{tabular}
\end{table}

\section{\textbf{Conclusion}}
\label{conclusion}
In this paper, we proposed a No-Reference Light Field image Quality assessment model based on Micro-Lens Image (LF-QMLI). We theoretically analyze the significance of MLI in LF-IQA and extract features for our evaluator. The model can effectively measure the angular consistency and spatial quality. The results show that LF-QMLI achieves state-of-the-art performance. In the future, we will consider more advanced features on the MLI to improve our model.

\section*{acknowledgment} 
This work was supported in part by NSFC under Grant 61571413, 61632001.

\footnotesize
\bibliographystyle{IEEEtran}
\bibliography{bibfile}

\end{document}